\documentclass[pre,twocolumn,amsmath,superscriptaddress,amssymb]{revtex4}
\usepackage[dvips]{graphicx}%include figures
\usepackage{dcolumn}%Align table columns on decimal point
\usepackage{wrapfig}

\begin{document}

\title{Dynamics of the fast component of nano-confined water under electric field}
\author{Souleymane Diallo}\email{omardiallos@ornl.gov}
\affiliation{Oak Ridge National Laboratory, Oak Ridge, TN 37831, USA}
\author{Eugene Mamontov}
\affiliation{Oak Ridge National Laboratory, Oak Ridge, TN 37831, USA}
\author{Andrey Podlesnyak}
\affiliation{Oak Ridge National Laboratory, Oak Ridge, TN 37831, USA}
\author{Georg Ehlers}
\affiliation{Oak Ridge National Laboratory, Oak Ridge, TN 37831, USA}
\author{Nobuo Wada}
\affiliation{Department of Physics, Nagoya University, Chikusa-ku, Nagoya 464-8602, Japan}
\author{Shinji Inagaki} % {S. Inagaki}
\author{Yoshiaki Fukushima} %{Y. Fukushima}
\affiliation{Toyota Central R\&D Laboratories, Inc., Yokomichi, Nagakute, Aichi 480-1192, Japan}

\begin{abstract}
We report the diffusion of water molecules confined in the pores of folded silica materials (FSM-12 with average pore diameter of $\sim$ 16 \AA), measured by means of quasielastic neutron scattering using the  cold neutron chopper spectrometer (CNCS). The goal is to investigate the effect of electric field on the previously observed fast component of nano-confined water. The measurements were taken at temperatures between 220 K and 245 K, and at two electric field values, 0 kV/mm and 2 kV/mm.  Similar to the recently observed electric field induced enhancement of the slow translational motion of confined water, there is a an equally important impact of the field on the faster diffusion.
\end{abstract}

\maketitle
\date{\today}

\section{Introduction}
The behavior of liquid water under a variety of conditions plays a fundamental role in many natural and technological processes. The effects of thermodynamic variables such as pressure and temperature, on the properties of water are of great fundamental interests, and for years have been investigated with neutron scattering \cite{Faraone:04,Liu:04, Liu:05, Takahara:05}. In contrast, studies of the behavior of water under other external parameters (such as electric field or magnetic field) while interesting are less common due largely to limitations in experimental design. Thanks to advances in sample cell design and in sample environment, some of these limitations can now be overcome, allowing new science to be explored. An important topic of interests in environmental science, biology, and industry concerns the ordering or crystallization of water molecules induced by the application of an external electric field for practical applications in cryopreservation of living cells and tissues, food processing, prevention of ice formation on crops or power lines and so on. Because water is a polar molecule, its electric dipole moment can be aligned in the direction of the applied electric field, yielding a more ordered structure. In a recent Science article, Ehre {\it et al.} \cite{Ehre:10} showed using simultaneously optical microscopy, calorimetry and X-ray diffraction that the temperature at which supercooled water freezes changes depending on whether the surface it rests on is positively or negatively charged or if it is neutral. Although such an effect has long been predicted, there is still no consensus on the actual microscopic picture \cite{Zhu:90,Vegiri:02} and its observation remains great experimental challenge. Early experiments relied on the use of metal plates (with distilled water sample) \cite{Vegiri:02} to generate the electric field, and were thus not able to isolate the sole effect of the field from those due the conducting plates, notably the fact that water molecules typically bind to highly conducting metallic surfaces near melting. Ehre {\it et al.} was able to screen these effects by using single crystals of the insulator LiTaO$_3$ coated with a SrTiO$_3$ film as electrodes. They found that a positively charged surface of this insulating electrode promotes water freezing by lowering its crystallization temperature down to 262 K, whereas the same plate charged negatively increases it to 265 K. These remarkable findings, which included structural studies, have naturally led us to want to investigate the effects of an external electric field on the dynamics of confined water.  Furthermore, a recent SEM report \cite{Choi:05} on interfacial water confined within a nanometer gap (14 \AA) has revealed that water molecules can order even at room temperature upon the application of a relatively small threshold electric field E$_c$. The reported E$_c$ is ~1 kV/mm, a value that is readily accessible in the laboratory as it is three orders of magnitude weaker than the field required for crystallization into polar cubic ice. Furthermore, molecular dynamics (MD) simulations of water confined in nanotubes at fixed temperature of 200 K \cite{Vegiri:02} have identified two remarkable critical fields (much larger than E$_c$), one due to the alignment of dipoles and the other to the onset of crystallization. These two critical fields are likely tied to the onset of different water dynamics. 

In a recent study using the high resolution backscattering instrument BASIS (which has a dynamics range of $\pm$0.1 meV),  we found a clear enhancement of the  translational diffusion of confined water when a moderate electric field of just 2.5 kV/mm is applied \cite{Diallo:12}.  These measurements revealed the existence of two diffusion components; a faster component which is   spatially restricted (or \lq caged' motion) and a slower unrestricted (or \lq uncaged') motion.  The application of an external electric field significantly increases the mobility of water molecules contributing to the slower process. But it was not clear whether the effect on the fast component was marginal or significant since this process is only partially observable on BASIS. We thus have re-investigated the effect of electric field using the  medium resolution Cold Chopper Spectrometer (CNCS) at the 1 MW Spallation Neutron Source (SNS), Oak Ridge National Laboratory (ORNL) \cite{Ehlers:11}, with the aim of investigating the effect of electric field on the fast component.  Our results indicate that there is indeed an increased observable quasi-elastic neutron scattering (QENS) broadening under field, which is associated with the fast diffusion process. This increased in mobility is of greater significance than it first appears to be on BASIS where the dynamics range is much limited. We conclude that investigations of  the broad component are more reliably done with medium resolution time-of-flight spectrometers such as CNCS, while accurate studies of the slow component require the use of high resolution backscattering instruments such as BASIS. 

\begin{figure}
\includegraphics[width=3.0in]{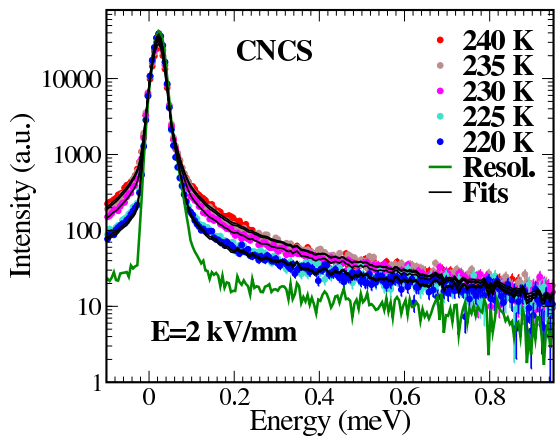}
\caption{ (Color) Temperature dependence of the quasielastic response of  water confined in FSM-12 (16 {\AA}) averaged over 0.5$\leq Q \leq$1.1 \AA$^{-1}$, as measured on CNCS. The solid symbols are the experimental data and the black lines are model fits, as described in the text. Resolution function measured with the same exact sample at 80 K is shown for comparison.}
\label{fig1}
\end{figure}

\section{Experimental details}
Water has been confined in the pores of  folded mesoporous insulating silica (FSM-12) by exposing the FSM-12 sample to a humid atmosphere at room temperature  until its mass increased by about 50\%. Prior to the hydration, the sample was dried for 48 hours under vacuum at 363 K. Our particular FSM-12 sample has an average pore diameter of 16 {\AA}. The synthesis of this highly ordered silica sample has been described elsewhere\cite{Inagaki:93}. The surface area (1130 m$^2$/g), the pore volume (0.51g/cc), and the pore diameter (16 \AA) were determined from N$_2$ adsorption isotherms.   The hydrated powder sample was transferred in specially designed Al flat cell for high voltage (HV) applications. The  HV cells consisted of two thin Al plates separated by Teflon gaskets. The effective gap between the plates was  $\sim$1.0 mm. The sample assemblies were sealed using a Viton O-ring to avoid contact between the plates. Two electrical leads, one attached to the positive port of a 10 kV TREK$^{TM}$ power supply and the other to the corresponding electrical ground, were connected to the sides of each HV cell, making the application of electric field possible. The Al sample cell was mounted in transmission geometry at 135$\deg$ with respect to the incident beam. Given our cell geometry and sample details, we estimated the loaded cell transmission to be about 94\% so that multiple scattering correction was not necessary.

The CNCS spectrometer was operated using an incident energy of 1.55 meV, which yielded a useful dynamics range for QENS of about $\pm$1meV. The water filled FSM-12 sample was placed in a flat Al-container and spectra were collected from 250 K down to 220 K. With the electric field, measurements were performed only between 220 K and 240 K, due to complications with the high voltage equipments. The resolution of the instrument was obtained by measuring the same Al flat which contained the FSM-12 sample at a temperature of ~80 K where we do not expect any mobility from the nano-confined water. A gaussian fit to this measured resolution integrated over $0.5<Q<1.1$ \AA$^{-1}$, yielded a  Full-Width at Half Maximum (or FWHM) of about 29 $\mu$eV at the elastic position. This resolution function is shown in Fig. \ref{fig1} along with QENS spectra collected at several temperatures for the 2 kV/mm electric field strength. As expected, the QENS signal becomes narrower as the temperature is reduced.

\section{Analysis and Results}
\begin{figure}
\includegraphics[width=3.0in]{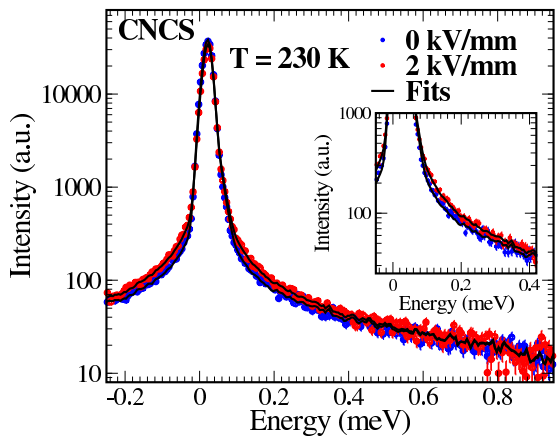}
\caption{(Color) Observed QENS response at a temperature of 230 K, at respectively zero electric field and 2 kV/mm. The data have been averaged over the wavevector $Q$, as described in the text. The solid symbols are the experimental data and the solid lines are model fits.  The inset shows that the QENS broadening due to the field is noticeably larger on a closer look. The impact of this broadening on the diffusivity of nano-confined water is of  comparable relevance to that observed on the slow component (see Fig. \ref{fig3}).}
\label{fig2}
\end{figure}

\begin{figure}
\includegraphics[width=3.2in]{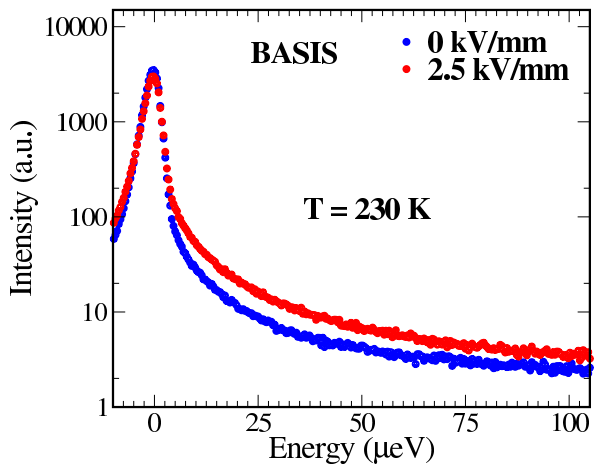}
\caption{(Color) Comparison between the QENS response at zero electric field and under moderate field averaged over $Q=0.3-0.9$ \AA$^{-1}$ and $T=$ 230 K(From Ref. [\onlinecite{Diallo:12}]). }
\label{fig3}
\end{figure}

\begin{table}
\caption{Temperature and field dependence of the fit parameters using Eq. \ref{stretch}. Parameters in the middle column have been rounded to the nearest decimal point.}
\begin{ruledtabular}
\begin{tabular}{| c  c | c c c | c |}
\newline
T (K) & Field  &  $\langle A\rangle_T$ &          $\tau_T$ (ps)  &     $ \beta_T$  & $ \langle \tau_a \rangle_T$ (ps)\\
\hline 
220   &off        &       0.86    &    31.97          &                0.34           & 179.4\\
          &{\bf on}        &       0.91    &     24.90          &               0.38   & {93.7} \\   
\hline
225   &off          &      0.85   &     30.96          &               0.34   &  163.3\\
          &{\bf on}         &      0.92   &     21.01          &               0.46   &{48.93} \\
\hline
230   &off         &       0.84    &    29.99            &             0.35   &  149.1\\ 
         &{\bf on}        &      0.82    &     21.11          &               0.51  & {40.1} \\
\hline
235   &off         &       0.83    &    28.18          &               0.37   & 114.0  \\
          &{\bf on}       &       0.76   &     20.78        &                 0.55   &  {35.3}\\
\hline
240   &off          &       0.83    &     21.90         &                0.38   & 83.1  \\
         &{\bf on}         &       0.73    &      20.77         &                0.55  &  {29.6}\\
\hline
245   &off          &       0.82    &   20.07          &               0.39     & 72.7\\
          &{\bf on}         &                                          &                      &       &     \\
\hline
250   &off          &        0.81  &    19.41            &             0.43      &52.59 \\
          &{\bf on}         &&& &\\
%\hline
%260   &off          &        0.75   &     17.88           &              0.44    & 44.3\\
%          &{\bf on}         &&& &\\
\hline
%270   &off          &        0.68   &    16.89        &                0.51      &32.58\\
%         &{\bf on}       &&&&
\end{tabular}
\end{ruledtabular}
\label{tbl1}
\end{table}

Previously \cite{Diallo:12}, we used a two Lorentzian model to capture all relevant motions in the data collected at BASIS at all temperatures. We thus anticipated to be able to describe the current data using a single Lorentzian function since the slow component is likely to be resolution limited on CNCS. On that assumption, we began our analysis  using a single Lorentzian model to capture the fast process only. Unfortunately, due to the limited statistical precision of the present data, the fits were rather poor, and the wavevector and temperature dependence of the fitting parameters were not consistent with each other, forcing us to abandon this approach. In the end, we resorted to using the  $Q$-integrated QENS spectra  to improve the statistics and the quality of the fits. The data reported have thus been summed over a $Q$ range 0.5$\leq Q \leq$1.1 \AA$^{-1}$  for each the temperature and field value quoted here. As a result of this $Q$-averaging, we use a stretched exponential model convoluted with the instrument resolution function to extract an average relaxation time for each spectra. The stretched function which gave good fits to the data can be written as \cite{Mansour:02,Takahara:05}, 
\begin{eqnarray} 
S_T(E)&=&I_0\big[\langle A\rangle_T\delta(E)+(1-\langle A\rangle_T)\times \label{stretch}\\ 
        &&\int_0^\infty dt~\cos(Et/\hbar)e^{-(t/\tau_T)^{\beta_T}} + a_TE+b_T\big] .    \nonumber
\end{eqnarray} 
\noindent The corresponding average characteristic relaxation time for such a function is given by, $\langle \tau_a\rangle_T=(\tau_T/\beta_T) \Gamma(1/\beta_T)$, where $\Gamma$ is the gamma function. There is a total of six fitting parameters in Eq. \ref{stretch}; the arbitrary intensity scale factor, the relative intensity of the elastic contribution $\langle A\rangle_T$, the relaxation time $\tau_T$, the stretching exponential $\beta_T$ and the two linear background terms, $a_T$ and $b_T$. A summary of the fit results using this model are reported in table \ref{tbl1}. The results indicate that; (a) independently of whether the electric field is applied or not, the water dynamics slow down as the temperature is reduced, yielding an average $\langle \tau_a\rangle_T$ that goes up with decreasing temperature and, (b) the water mobility is increased by the application of field, as evidenced by the relative QENS broadening. The fast diffusion component is clearly affected by the application of electric field at all temperatures. This effect appears to be larger than previously suggested by the BASIS data alone where the broad component was only partially captured.

\begin{figure}
\includegraphics[width=3.2in,angle=-90]{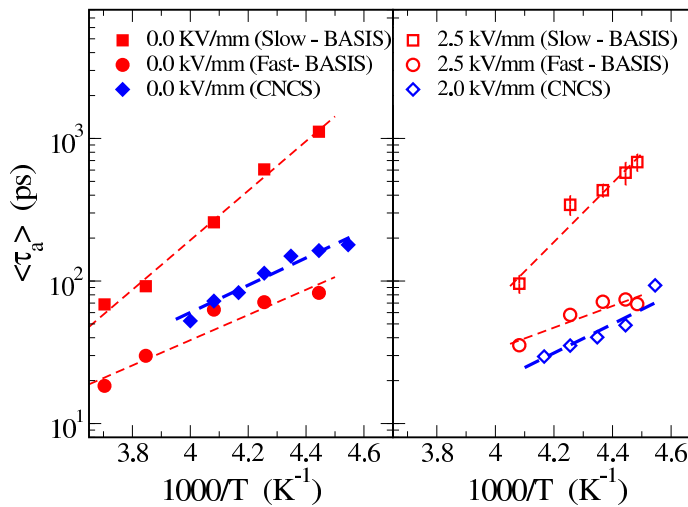}
\caption{(Color) Temperature dependence of the observed relaxation times associated with the diffusivity of water confined in nano-porous silica FSM-12, as measured on the backscattering instrument BASIS and the time-of-flight spectrometer CNCS.  The solid symbols represent the zero field data (left panel) and the open symbols the applied field data (right panel). The different components are discussed in the text.}
\label{fig4}
\end{figure}

Fig. \ref{fig2} compares the QENS broadenings at temperature $T=230$ K with and without electric field, observed on CNCS. When a field $E= 2$ kV/mm is applied across the sample, the QENS signal becomes broader. This is in agreement with our earlier observation on the slow diffusion component (see Fig. \ref{fig3}), and suggests that the fast motion of water molecules is also enhanced with moderate field values. The observed increased in water mobility is of significant relevance as that observed on the slow component (see inset of Fig. \ref{fig2} and Fig. \ref{fig3}). This  field enhanced mobility on the fast component is better illustrated in Fig. \ref{fig4} which shows the temperature dependence of the observed relaxation times of water confined in FSM-12. The figure summarizes our present measurements compared with those in Ref. \cite{Diallo:12}.  The results in the absence of field are shown in the left panel and those in the presence of field in the right panel. The field induced response of the fast component  is likely due  to the disruption of the hydrogen bonding, which arises from a competition between the tetrahedral coordination due to the hydrogen bonds and the dipolar alignments due to the field. In this event,  the mobility of individual water molecules is enhanced due to disorder. The system eventually reverts back to a relatively more ordered state either when the field is switched off or when the majority of the dipoles are aligned along the field. In the latter scenario, we anticipate a slowing down of the water molecules due to a lower entropy. Independent molecular dynamics simulations and further measurements at much higher field strength will help test this hypothesis.

In the analyzed temperature range, the relaxation time can be fit to an Arrhenius temperature dependence $\langle\tau_a\rangle=\tau_{0}e^{\frac{E_{A}}{RT}}$,  where $E_A$ is the activation energy. The Arrhenius fit parameters are summarized in Table \ref{tbl2}, and the corresponding fits are shown as dashed lines in Fig. \ref{fig4}. The parameters $\tau_{0}$ is multiplicative pre-factors associated with the faster vibrational processes, not observable in our current dynamics range. The parameters $\tau_0$ and $\langle\tau_a\rangle$ denote respectively the intercept  and the slope of the fitted lines. From inspection of Fig. \ref{fig4}, we note that the slow and fast components are well separated in time, by 1-2 orders of magnitude with or without the field applied and thus very distinguishable. However, the slow component is more reliably measured on BASIS while the fast component is best investigated with CNCS. This explains why the effect of field on the fast component  appears to be less significant on BASIS than it really is.

\begin{table}
\caption{Arrhenius fit  parameters of the average relaxation time $\langle \tau_a\rangle_T$ in the absence and presence of electric field.}
\begin{ruledtabular}
\begin{tabular}{|c|c|c|}
 $E$ (kV/mm)                 & $\tau_{0}$ (ps)   & E$_{A}$ (kJ/mol) \\
\hline
          0                     &0.008 &  18.32  \\
 \hline 
         2                & 0.016  &  19.50   \\
\end{tabular}
\end{ruledtabular}
\label{tbl2}
\end{table}

%\begin{figure}
%\includegraphics[width=22.0in]{Tau_plots.eps}
%\caption{(Color) Temperature dependence of the average relaxation time $\langle\tau_a\rangle_T$,  of the fast component of water in FSM-12, as measured on CNCS. The red squares represent the $E=0$ kV/mm data and the black squares  the $E=$ 2 kV/mm data.  The reported  $\langle\tau_a\rangle_T$ were determined at each temperature from fitting Eq. \ref{stretch} to data averaged over $Q$ range from 0.5 to 1.1 \AA$^{-1}$ at that temperature. }
%\label{fig4}
%\end{figure}

\section{Conclusions}
We have used quasi-elastic neutron scattering to probe the effect of moderate electric field on the fast component of water diffusing in a regular array of pores in FSM-12. Generally, our results confirm that the translation diffusion of water is enhanced in the presence of a moderate electric field. Using a medium resolution of 29 $\mu$eV, we find an observably broader QENS signal, mostly noticeable at low energies, in the presence of electric field. Future studies which would include information on the wavevector dependence of the relaxation processes could probe geometrical information and provide additional details. Understanding the enhancing effect of electric field on nano-confined water and testing the universality of this behavior in other silica substrates or biological relevant systems such as Lysozyme is of great scientific importance. In this endeavor, additional measurements and detailed molecular dynamics studies of nano-confined water under moderate electric field  would have to be performed.

Thanks are due to S. Elorfi, D. Maierhafer, R. Mills, and M. Rennich  at SNS for valuable technical support. We acknowledge the valuable use of the DAVE software package \cite{Azuah:09}. Work at ORNL and SNS is sponsored by the Scientific User Facilities Division, Office of Basic Energy Sciences, US Department of Energy.

%{\footnotesize
%\begin{thebibliography}{26}
%\end{thebiobligraphy}
%}

%{\footnotesize

%\bibliographystyle{unsrt}
%\bibliography{bibs/abbrevs,bibs/water,bibs/water_A-F,bibs/water_G-P,bibs/water_R-Z}
%}
{\footnotesize

\end{document}